\documentclass[10pt,aps,pra,twocolumn,superscriptaddress]{revtex4-1}

\usepackage{graphicx,braket,amsmath}
\usepackage[margin=0.75in]{geometry}

\begin{document}

\title{Efficiency of an enhanced linear optical Bell-state measurement scheme with realistic
imperfections}

\author{Stephen Wein}
\affiliation{Institute for Quantum Science and Technology and
Department of Physics and Astronomy, University of Calgary, Calgary, Alberta,
Canada T2N 1N4}

\author{Khabat Heshami}
\affiliation{Institute for Quantum Science and Technology and
Department of Physics and Astronomy, University of Calgary, Calgary, Alberta,
Canada T2N 1N4}
\affiliation{National Research Council of Canada, 100 Sussex Drive, Ottawa, Ontario, Canada K1A 0R6}

\author{Christopher A. Fuchs}
\email{Present affiliation: University of Massachusetts Boston, 100 Morrissey
Boulevard, Boston, MA 02125, USA }
\author{\mbox{Hari Krovi}}
\author{Zachary Dutton}
\affiliation{Quantum Information Processing Group,
Raytheon BBN Technologies, Cambridge, USA}

\author{Wolfgang Tittel}
\author{Christoph Simon}
\affiliation{Institute for Quantum Science and Technology and Department of
Physics and Astronomy, University of Calgary, Calgary, Alberta, Canada T2N 1N4}

\begin{abstract}
We compare the standard 50\%-efficient single beam splitter method for Bell-state
measurement to a proposed 75\%-efficient auxiliary-photon-enhanced scheme [W. P. Grice, Phys. Rev. A \textbf{84}, 042331 (2011)] in light of realistic conditions. The two schemes are compared with consideration for high input state photon loss, auxiliary state photon loss, detector inefficiency and coupling loss, detector dark counts, and non-number-resolving detectors. We also analyze the two schemes when multiplexed arrays of non-number-resolving detectors are used. Furthermore, we explore the possibility of utilizing spontaneous parametric down-conversion as the auxiliary photon pair source required by the enhanced scheme. In these different cases, we determine the bounds on the detector parameters at which the enhanced scheme becomes superior to the standard scheme and describe the impact of the different imperfections on measurement success rate and discrimination fidelity. This is done using a combination of numeric and analytic techniques. For many of the cases discussed, the size of the Hilbert space and the number of measurement outcomes can be very large, which makes direct numerical solutions computationally costly. To alleviate this problem, all of our numerical computations are performed using pure states. This requires tracking the loss modes until measurement and treating dark counts as variations on measurement outcomes rather than modifications to the state itself. In addition, we provide approximate analytic expressions that illustrate the effect of different imperfections on the Bell-state analyzer quality.
\end{abstract}
\date{\today}

\maketitle

\section{Introduction}
\label{sec:intro}
\vspace{-1mm}

Bell-state measurements are a key component of quantum information processing protocols such as quantum teleportation \cite{teleportation}, entanglement swapping \cite{zukowski1993} and dense coding \cite{dense}. They are particularly critical in the context of quantum communication. The implementation of large-scale quantum networks for long-distance quantum cryptography \cite{gisin2002} and a future quantum internet \cite{kimble2008} requires the distribution of photon entanglement over long distances. Unfortunately, direct photon transmission through fibers suffers from exponential loss over distance causing rates to decrease below acceptable levels \cite{gisin2002,
takeoka2014, pirandola2015}. In classical communication this problem is circumvented by signal
amplification. However in the quantum world, this cannot be accomplished due to
the no-cloning theorem for quantum states \cite{dieks1982, wootters1982}.

Many promising methods that overcome the direct transmission loss-rate trade-off
use quantum repeaters
\cite{briegel1998,duan2001,simon2007,jiang2009,sangouard2011,
sinclair2014,guha2014,krovi2015}.
These methods rely on generating entanglement over multiple shorter distances, and
then connecting the links by entanglement swapping \cite{zukowski1993},
to extend the entanglement over the entire distance. Free-space
based proposals using satellites take advantage of reduced photon absorption
outside the atmosphere, allowing further distances
\cite{buttler1998,bonato2009,wang2013,bourgoin2014,boone2015}. However, global entanglement distribution using satellites would also require entanglement swapping due to earth curvature
\cite{boone2015}. An ideal distribution network may comprise a combination
of fiber and free-space solutions, using fibers for regional distribution and
satellites for intercontinental distribution.

A shared requirement among all practical entanglement distribution protocols is
successful entanglement swapping between neighboring nodes to create links.
This is done by applying a projective Bell-state measurement (BSM) on two
photons, one from each node \cite{gisin2002}. The measurement result would then
indicate the entangled state of the link. Ideally, each measurement outcome should
unambiguously indicate projection onto one of the four Bell states. Unfortunately, a perfect Bell-state analyzer cannot be
constructed using only linear optical elements \cite{lutkenhaus1999,vaidman1999} and cannot exceed 50\% success rate without the addition of auxiliary photons \cite{lutkenhaus2001}. The simplest
Bell-state analyzer can resolve at most two of the four states resulting in a maximum
50\% success rate for entanglement swapping \cite{braunstein1995}. This is a
source of inefficiency in all repeater schemes and compounds with an increase in
the number of swapping events.

A perfectly complete
Bell-state analyzer cannot be realized by simple means; however, efficiency can be increased
arbitrarily close to unity at the cost of increased apparatus complexity and
added auxiliary photons \cite{grice2011}. As a consequence of
increasing complexity, the design can become more susceptible to component flaws
that reduce efficiency and fidelity, offsetting initial gains.
Hence it is necessary to compare BSM schemes in light of imperfections such as non-number-resolving detectors, detector inefficiencies, dark counts, and photon loss rates.

To approach the problem of quantifying the analyzer quality, we use a combination of numeric and analytic techniques. Determining the exact quality of a Bell-state analyzer, when accounting for different imperfections or auxiliary sources, can be analytically difficult and computationally costly. This is because the size of the Hilbert space and the number of measurement outcomes increases quickly when additional parameters, detectors, and photons are considered. Our approach to this problem is to use only pure states when performing numerical computations, which reduces the required computational resources. This is accomplished by tracking all loss modes until the measurement is performed and by treating dark counts as variations on measurement outcomes rather than altering the output state. We also perform analytic approximations to derive expressions that give insight into the influence of imperfections on the Bell-state analyzer quality and use our numerical results to verify the accuracy of these expressions.

This paper is organized as follows. Section \ref{sec:back} introduces the Bell measurement schemes, detector type models, loss considerations, and the alternative
auxiliary photon source used in our analysis. Section \ref{sec:meth} outlines the methods used to analyze the schemes under realistic
conditions, including simple examples using the standard scheme as a case
study. Numeric and analytic results are reported in section
\ref{sec:results}, which is divided into three parts. The first part explores
replacing detectors with imperfect ones, the second
considers multiplexed arrays of non-number-resolving detectors, and the third
deals with an alternative auxiliary source. Finally, we summarize our conclusions in section \ref{sec:conc}.

\section{Background and definitions}
\label{sec:back}

\subsection{Bell-state measurement schemes}
\label{ssec:bms}

A Bell-state analyzer is a device that performs a (partial) Bell-state measurement. It
is characterized by its success rate for projecting an entangled state onto one
of the four Bell states and identifying the result, assuming a uniform
distribution of all four Bell states as input. These Bell states can be written
as
\begin{equation}
\begin{aligned}
\ket{\psi^\pm}_{12} &=
\frac{1}{\sqrt{2}}\left(\hat{a}^{\dagger}_1\hat{b}^{\dagger}_2\pm
\hat{b}^{\dagger}_1\hat{a}^{\dagger}_2\right)\ket{0}_{12}\\
\ket{\phi^\pm}_{12} &=
\frac{1}{\sqrt{2}}\left(\hat{a}^{\dagger}_1\hat{a}^{\dagger}_2\pm
\hat{b}^{\dagger}_1\hat{b}^{\dagger}_2\right)\ket{0}_{12}\;,
\end{aligned}
\end{equation}
\noindent
where $\hat{a}^{\dagger}$ and $\hat{b}^{\dagger}$ are photon creation operators
for qubit mode and the subscript denotes spatial mode.

\begin{figure}[t]
\includegraphics[width=8.5cm]{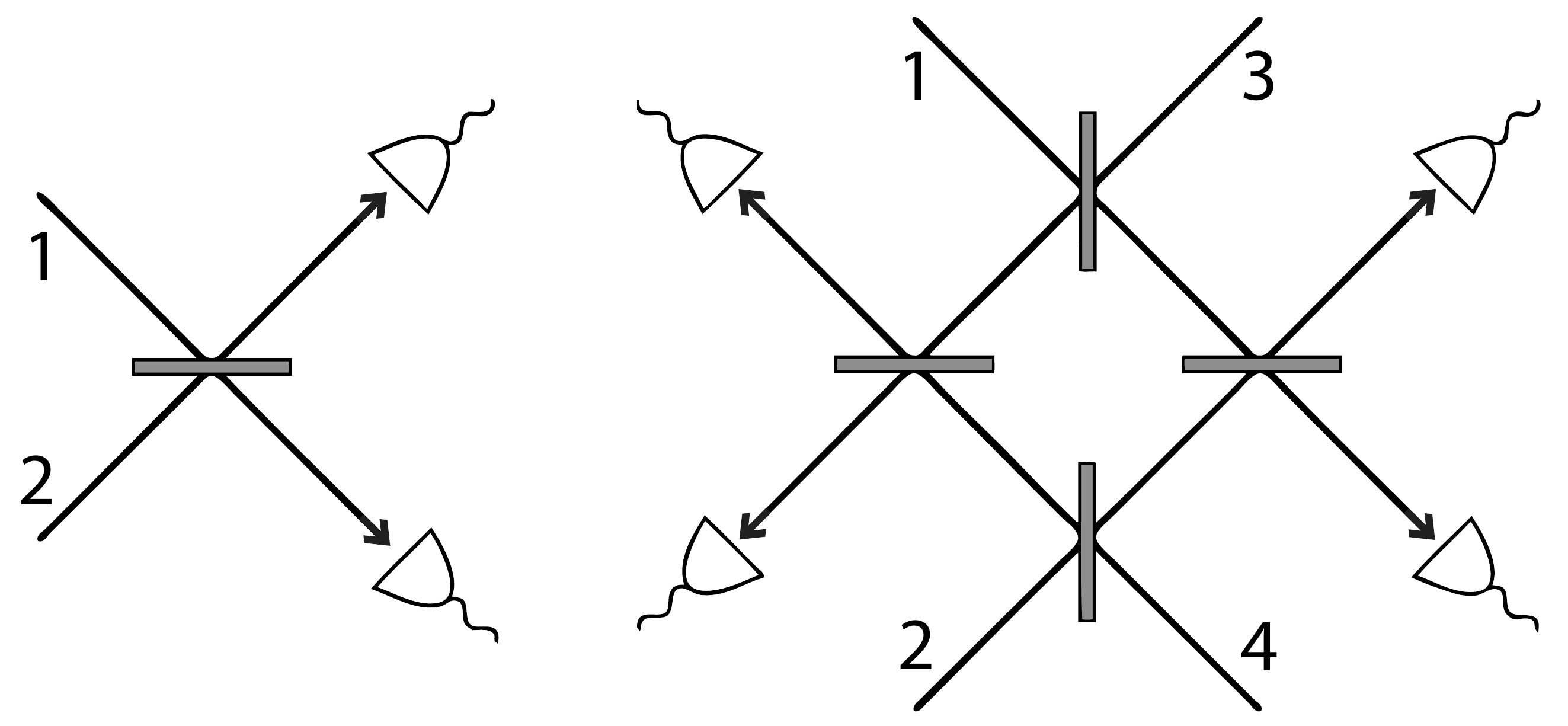}
\caption{
(left) A schematic of the standard method for performing a Bell-state
measurement.
(right) Grice's proposed first-order extension for Bell-state measurement.
The spatial modes $1$ and $2$ represent the input pair while the spatial modes
$3$ and $4$ represent the auxiliary pair. Gray bars represent non-polarizing
beam splitters. Diagrams are shown assuming measurement of time-bin
encoded qubits. For polarization encoded qubits, each detector in the above
diagrams is replaced with a polarizing beam-splitter and two detectors. For
dual-rail encoded qubits, the above diagrams are duplicated, one for each rail.}
\label{fig:schemediagrams}
\end{figure}

The most standard 50\%-efficient scheme for conducting a BSM is composed of a
single non-polarizing beam splitter and two detectors \cite{braunstein1995},
also referred to as a Hong-Ou-Mandel interferometer, as illustrated in FIG. \ref{fig:schemediagrams} (left). This is described by the beam splitter transformation
\begin{equation}
\label{beamsplitterT}
\begin{pmatrix}
\hat{c}^{\dagger}_1\\
\hat{c}^{\dagger}_2
\end{pmatrix}_{\text{in}}
\rightarrow
\frac{1}{\sqrt{2}}
\begin{pmatrix}
\ 1\ &\ i\ \\
i&1
\end{pmatrix}
\begin{pmatrix}
\hat{c}^{\dagger}_1\\
\hat{c}^{\dagger}_2
\end{pmatrix}_{\text{out}}\;,
\end{equation}
\noindent
where $\hat{c}^\dagger$ is replaced by $\hat{a}^\dagger$ or $\hat{b}^\dagger$ as appropriate. This Bell-state analyzer has at
most a 50\% success rate since only $\ket{\psi^+}$ and $\ket{\psi^-}$ of the
four Bell states are distinguishable by measurement outcomes. The other two Bell states
$\ket{\phi^\pm}$ cannot be distinguished from each other but can be
distinguished from $\ket{\psi^\pm}$. The simple design and lack of auxiliary
photons makes this method attractive for practical implementations. However,
limited success rate is undesirable especially when many measurements are
required to generate entanglement over long distances.

We compare this standard scheme Bell-state analyzer to an extension proposed by W. P. Grice \cite{grice2011}. In his paper, Grice defined a series of enhanced Bell-state analyzers using increasingly complicated entangled auxiliary states that, in the series limit, show an arbitrarily complete Bell measurement. The auxiliary states required to approach a perfect Bell-state analyzer are impractical to construct; however, the first-order extension only requires an auxiliary $\ket{\phi^+}_{34}$ Bell state to achieve a 75\%-efficient Bell measurement, which is the enhanced scheme studied in this paper. This first-order extension utilizes four beam splitters and four
detectors, as illustrated in FIG. \ref{fig:schemediagrams} (right), and is described by the transformation
\begin{equation}
\begin{pmatrix}
\hat{c}^{\dagger}_1\\
\hat{c}^{\dagger}_2\\
\hat{c}^{\dagger}_3\\
\hat{c}^{\dagger}_4
\end{pmatrix}_{\text{in}}
\rightarrow
\frac{1}{2}
\begin{pmatrix}
\ 1\  &\ i\  &\ i\  &\ -1\ \\
\ i\  &\ 1\  &\ -1\ &\ i\  \\
\ i\  &\ -1\ &\ 1\  &\ i\  \\
\ -1\ &\ i\  &\ i\  &\ 1\
\end{pmatrix}
\begin{pmatrix}
\hat{c}^{\dagger}_1\\
\hat{c}^{\dagger}_2\\
\hat{c}^{\dagger}_3\\
\hat{c}^{\dagger}_4
\end{pmatrix}_{\text{out}}\;,
\end{equation}
\noindent
where spatial modes 1 and 2 contain the input state to be measured and spatial modes 3 and 4 contain the required auxiliary state.

\noindent

\subsection{Detector types}
\label{ssec:dettypes}

Driven by the rapid expansion of quantum information science, single-photon
detectors (SPDs) have seen significant improvements over the past few decades
\cite{hadfield2009,eisaman2011}. In particular, photon-number-resolving
detectors (PNRDs) that use transition-edge sensors (TESs) have reached excellent
efficiencies of 95\% at telecommunication wavelengths
\cite{rosenberg2005,lita2008}. Unfortunately TES-based detectors operate at
temperatures of 0.1 K and also have recovery time on the order of microseconds
\cite{eisaman2011}. Long detector recovery time or `dead time' is particularly
undesired for applications that use time-bin encoded qubits, which are popular
in quantum communication
\cite{tittel1998,tittel1998a,brendel1999,tittel2000,sinclair2014,valivarthi2014}.
Faster detectors such as single-photon avalanche diode (SPAD) detectors and
recently developed superconducting nanowire SPDs (SNSPDs) have recovery times on
the order of nanoseconds \cite{hadfield2009,eisaman2011,marsili2013}, which are
more suited for use with time-bin qubits. Although they operate at higher
temperatures (77 K - 250 K), SPAD detectors are relatively inefficient for
telecommunication wavelengths ($\sim$ 10\%) and cannot natively resolve photon
number \cite{hadfield2009,eisaman2011}. On the other hand, SNSPDs show
impressive efficiencies of 93\% for telecommunication wavelengths
\cite{natarajan2012} but, like SPADs, they cannot resolve photon number and, like
TESs, require significant cooling ($\lesssim$ 3 K) \cite{eisaman2011}. Currently
there does not exist a detector type that satisfies all desired criteria for
quantum communication applications; therefore careful analysis of the trade-offs
is required on a per-application basis. In this paper we discuss and analyze
some of these trade-offs in the context of a Bell-state analyzer.

We consider four different types of abstract detector models,
two of which are relevant to time-bin encoded applications only. The first and most ideal
detector model is a PNRD. In addition to resolving photon number, we assume a
PNRD has negligible dead time. Secondly we consider a non-number-resolving SPD
that cannot indicate photon number but again has negligible dead time. That is,
it is a binary detector (BD) that clicks when photons are incident and does not
click when photons are not incident, similar to a SPAD detector or an SNSPD
\cite{hadfield2009,eisaman2011}.
We assume for simplicity that both the PNRD and BD can perfectly resolve qubit
modes. This can be implemented, for example, by placing a polarizing beam
splitter prior to the detectors (for polarization encoded qubits), by placing a
detector on each spatial rail (for dual-rail encoded qubits), or recording two
time-separated events (for time-bin encoded qubits).

The last two detectors we considered are similar to the PNRD and BD described
above but have the additional flaw of long dead time. This flaw is only relevant
to time-bin-encoded qubits provided the temporal modes are restricted to a
maximum temporal separation. We describe a slow PNRD that can indicate photon
number but has a dead time longer than the time separation between qubit modes,
which could occur with a TES detector \cite{rosenberg2005,lita2008}. As a
result, if one or more early photons are detected, the detector is unable to
count possible late photons in the same spatial mode. A slow BD would be similar
to a slow PNRD but again cannot resolve photon number, similar to a gated SPAD
\cite{hadfield2009,eisaman2011}. To give an example, a slow PNRD measuring a
state with two early photons and two late photons would indicate the presence of
two early photons whereas a slow BD measuring the same state would indicate only
one early photon. In addition to number-resolving imperfections and long recovery times, each detector model is assumed to also suffer from detection inefficiencies and dark counts.

Alongside the development of PNRDs such as the TES, detectors that attempt to
resolve photon number by multiplexing many non-PNRDs in either space
\cite{divochiy2008, jiang2007} or time \cite{achilles2003} are also being
pursued. These methods attempt to combine the benefits of non-PNRDs such as
warmer operation temperature or short dead times with number resolving
capabilities. These multiplexed detectors can be modeled by arrays of imperfect
detectors; however, undesirable effects caused by inefficiency and dark counts
can be amplified by large detector arrays. We discuss regimes in which detector
arrays of BDs and slow BDs can increase Bell-state analyzer success rates closer to those obtainable with PNRDs.

\subsection{Photon loss channels}
\label{ssec:loss}

The input state that is to be measured can suffer from photon loss. In particular, when utilizing the Bell-state analyzer in repeater elementary links, the input photons must first travel large distances. As a consequence, the input state to the Bell-state analyzer can have poor transmission rates \cite{krovi2015}. Aside from a reduction in total efficiency of the measurement, this input loss can affect the fidelity of the scheme when considered along with detector dark counts.

For the enhanced scheme, the auxiliary state can also suffer from photon loss. This loss channel represents a contribution from propagation absorption, coupling losses, and pair generation inefficiency. For bright auxiliary pair sources, this loss rate is expected to be small as compared to the input state photon loss rate. However, auxiliary state photon loss only affects the quality of the enhanced scheme, and not the standard scheme since it does not require auxiliary photons, and so this parameter is expected to directly affect the bounds on the regime where the enhanced scheme is superior to the standard scheme.

Finally, both schemes are also subject to photon loss between the beam splitters and the detectors. This loss rate can be accounted for as combined with detector efficiency to form an effective scheme loss/inefficiency parameter.

\subsection{Alternative auxiliary source}
\label{ssec:aux}

Schemes that improve success rates beyond 50\% require auxiliary photon states. These states are susceptible to realistic imperfections. In some cases, using approximations for these auxiliary states may also be necessary for implementation to
remain practical. The enhanced scheme studied in this paper requires an
auxiliary photon pair in a maximally entangled Bell state to achieve 75\%
success rate \cite{grice2011}. Although methods for producing single entangled pairs, such as biexciton decay in a quantum dots \cite{dousse2010}, are becoming more efficient, spontaneous parametric down-conversion (SPDC) remains the most widely available source of photon entanglement.

In addition to studying accurate auxiliary states, we also analyze the case where the auxiliary state is produced by a SPDC source. In this case, the $\ket{\phi^+}$ auxiliary Bell state can be approximated by pumping a nonlinear crystal to produce a multi-photon entangled state \cite{kwiat1995,simon2003,eisenberg2004}. The auxiliary state would then be given by
\begin{equation}
\label{pdctau}
\begin{aligned}
\ket{\Phi^+}_{34} &=\sum_{n=0}^\infty w(n,\tau)\ket{\phi_{n}^+}_{34}\\
w(n,\tau) &= \sqrt{n+1}\text{ sech}^2 \tau\text{
tanh}^n\tau\\
\ket{\phi^+_n}_{34}&=
\frac{1}{n!\sqrt{n+1}}(\hat{a}^{\dagger}_3\hat{a}^{\dagger}_4+
\hat{b}^{\dagger}_3\hat{b}^{\dagger}_4)^n\ket{0}_{34}\;,
\end{aligned}
\end{equation}
\noindent
where the parameter $\tau$ is the SPDC interaction parameter and is related to
mean photon number of the states being generated. The index $n$ here represents
the pair number so that terms $\ket{\phi^+_n}$ are entangled photon states with
$2n$ photons.

\section{Methods}
\label{sec:meth}

\subsection{Scheme quality: definitions and analysis}
\label{ssec:quality}

An ideal destructive BSM scheme must unambiguously distinguish each of
the Bell states without prior information on the state.
The quality of a BSM scheme is described by the measurement success rate and fidelity.

The maximum total success rate of a Bell-state analyzer is defined as the unweighted
average of the probabilities of unambiguously identifying each Bell state
individually \cite{grice2011,ewert2014}. This standard definition for Bell-state analyzer success rate requires the input to be in the span of the four Bell states, which form a complete basis for the bi-photon Hilbert space. That is, by standard definitions, a Bell-state analyzer performs a measurement on a bi-photon state. Therefore, this method gives an accurate representation of a Bell-state analyzer's performance if the input is well-approximated by a bi-photon state, otherwise the input is not in the span of the Bell basis and the Bell-state analyzer would no longer be projecting states onto just the Bell basis. We aim to provide a comparison of the two Bell-state analyzers that is relatively independent of the exact input photon state; hence, input states containing more than two photons are not treated in this work. These cases complicate the standard definitions of a Bell-state analyzer, often requiring application-specific analysis to compute success rates.

The maximum total success rate represents a quantum-limited upper bound on the efficiency of a Bell-state analyzer. The actual success rate of the measurement is less than the
maximum because it depends on the choice of measurement
outcomes used to identify the state and includes effects caused by
imperfections in equipment. To illustrate this point, we consider the
standard scheme. Applying the beam splitter
transformation given by Eq. (\ref{beamsplitterT}) to each Bell state results in the output states
\begin{equation}
\label{eq:outputs}
\vspace{-1mm}
\begin{aligned}
\ket{\psi^+} &\rightarrow
\frac{i}{\sqrt{2}}\left(\hat{a}^{\dagger}_1\hat{b}^{\dagger}_1+
\hat{a}^{\dagger}_2\hat{b}^{\dagger}_2\right)\ket{0}\\
\ket{\psi^-} &\rightarrow
\frac{1}{\sqrt{2}}\left(\hat{a}^{\dagger}_1\hat{b}^{\dagger}_2-
\hat{a}^{\dagger}_2\hat{b}^{\dagger}_1\right)\ket{0}
\\
\ket{\phi^\pm} &\rightarrow
\frac{1}{2}\left(\hat{a}^{\dagger 2}_1+\hat{a}^{\dagger
2}_2\pm\hat{b}^{\dagger 2}_1\pm\hat{b}^{\dagger 2}_2\right)\ket{0}\;,
\end{aligned}
\end{equation}
\noindent
where $\hat{c}^{\dagger}$ was replaced with $\hat{a}^{\dagger}$ or
$\hat{b}^{\dagger}$ appropriately. For a perfect PNRD detector model, each
output state is faithfully reported by the detector and so we are able to
distinguish outputs of $\ket{\psi^\pm}$ from each other and from the outputs of
$\ket{\phi^\pm}$. However, both the $\ket{\phi^\pm}$ outputs are identical but
for phase, thus the detector cannot distinguish between them.
Since only half of the Bell states are uniquely identified by measurement
outcomes, we have a scheme maximum success rate of $50\%$.

The BD model cannot resolve photon number and reports states
${\hat{a}^{\dagger 2}}_i\ket{0}$ and $\hat{b}^{\dagger 2}_i\ket{0}$ as
$\hat{a}^{\dagger}_i\ket{0}$ and $\hat{b}^{\dagger}_i\ket{0}$ respectively.
Fortunately these are still distinguishable from the $\ket{\psi^\pm}$ outputs,
again giving 50\%.

Our slow PNRD model reports states
$\hat{a}^{\dagger}_i\hat{b}^{\dagger}_i\ket{0}$ as $\hat{a}^{\dagger}_i\ket{0}$
due to photons created by $\hat{b}^{\dagger}_i$ arriving during detector dead
time. This change only affects the outputs of $\ket{\psi^+}$, causing them to
appear as one-photon events. Luckily, number resolution allows the detector to
faithfully represent the two-photon outputs of $\ket{\phi^\pm}$ and thus
$\ket{\psi^+}$ can still be distinguished from $\ket{\phi^\pm}$ based on photon
count. Since the outputs of $\ket{\psi^-}$ are also unaffected, this detector
model still allows for a 50\% maximum success rate.

The slow BD model, however, cannot faithfully resolve two-photon states and
hence cannot distinguish between output states of $\ket{\psi^+}$ and
$\ket{\phi^\pm}$, resulting in a detector-limited 25\% maximum success rate. This 25\%-efficient Bell measurement is often performed using time-bin-encoded photons if fast detectors are unavailable \cite{marcikic2003, riedmatten2005,jin2013}.

The enhanced scheme can be analyzed in a similar manner as the standard scheme but, for brevity, this will not be shown since the
$\ket{\psi^\pm}$ Bell states have 80 possible outputs each, $\ket{\phi^+}$ has
42, and $\ket{\phi^-}$ has 38. The unambiguous output states of the enhanced scheme include four-mode, three-mode, and two-mode occupied states. All single-mode states, where all four photons occupy the same mode, are ambiguous and must be rejected. Since there exists four-mode output states, where all photons occupy different modes, the success rate is nonzero when using non-number resolving detectors. The success rates for the enhanced scheme are given in section \ref{ssec:rates}.

In addition to reduced success rates, including imperfections in the analysis
can lead to a reduction in the measurement discrimination fidelity of the scheme. To quantify this, we organize the measurements into two mutually exclusive
categories: the positive (post-selected) measurements, and the negative (rejected) measurements. Within the positive measurements, a result can either be true (successful) or false (unsuccessful).
Hence, true positive measurements are those that collapse the state into a Bell state and correctly identify which Bell state it collapsed into. False positive measurements are those that are post-selected but either
do not collapse the state into a Bell state or falsely identify which Bell state it collapsed into. In this way, the true success rate of a BSM can be described by the
true-positive rate, denoted here by $p_t$. If we denote the false-positive rate by $p_f$, the total probability of a positive measurement is $p_t + p_f$ and the fidelity $f$ of a BSM can then be computed by
\begin{equation}
f=\frac{p_t}{p_t+p_f}
\end{equation}

\subsection{Dealing with large system sizes}
\label{ssec:sizes}

For a small apparatus like the standard scheme, measurement outcomes and probabilities can be computed by hand. However, when
enhanced schemes with additional auxiliary states are analyzed, the Hilbert space
can become very large and numerical methods become necessary. The size of an
$n$-photon Hilbert space with $k$ states per photon is
\begin{equation}
\label{hsize}
N_H=\binom{k+n-1}{n} = \frac{(k+n-1)!}{(k-1)! n!}\;.
\end{equation}
For example, the size of the Hilbert space for the standard scheme ($n=2$ and
$k=4$) is 10. However, the 4-detector enhanced scheme ($n=4$ and $k=8$) has a
Hilbert space size 330. Furthermore, if input and auxiliary loss is accounted for, then the incident
states can consist of zero, one, or two photons. In this case the size of the
Hilbert space for the standard scheme increases to 15 whereas for the enhanced
scheme, it increases to 495.

For number-resolving detector types, the number of measurement outcomes can be much larger than the Hilbert space size when including dark counts. For the enhanced scheme with
PNRDs including detector inefficiencies, dark counts, and two auxiliary photons, the number of
measurement outcomes is already 23392. The size of the Hilbert space exponentially
increases when additional resources are added, with SPDC cases and some detector
array numerical calculations in this study reaching Hilbert space sizes on the
order of millions. For example, the largest auxiliary state we compute contains 20 photons, resulting in a Hilbert space size of $N_H=1560780$ as determined using Eq. (\ref{hsize}) with $k=8$ and $n=22$.

Approaching these large computations using standard positive-operator valued measure (POVM) methods with the density operator formalism would require numerical arrays of sizes $N_H^2$ that can reach magnitudes of $\sim 10^{12}$ in some cases, which is impractically large for numerical methods. In addition, since the number of measurement outcomes can exceed the Hilbert space size, POVM operators can contain more than $N_H^2$ elements. For this reason, we approach the numerical problem by representing states as pure states rather than mixed states. This requires tracking the loss modes until measurement and treating dark counts as variations on measurement outcomes rather than modifications to the state itself. We expand our discussion on these methods in the following section.

\subsection{Loss, detector inefficiency, and dark counts}

To account for input photon loss prior to BSM, each Bell state was first passed
through a beam splitter with weight $\sqrt{\eta_i}$ for input state transmission rate
$\eta_i$. This transformation is given by
\begin{equation}
\label{BSloss}
\hat{c}^{\dagger} \rightarrow \sqrt{\eta_i}\hspace{1mm}\hat{c}^{\dagger} +
\sqrt{1-\eta_i}\hspace{1mm}\hat{l}^\dagger\;,
\vspace{-1mm}
\end{equation}
\noindent
where $\hat{l}^\dagger$ is the photon creation operator for the loss mode. For the enhanced scheme, we applied an identical transformation on the auxiliary state using transmission rate parameter $\eta_a$. This
causes the output state probabilities to become dependent on $\eta_i$ and also $\eta_a$ for the enhanced scheme.

To account for detector inefficiency, dark counts, and different detector models, we determined the
conditional probabilities $P(m_i|\varphi_j)$ that a measurement outcome $m_i$ is
triggered by a state $\ket{\varphi_j}$, where $\ket{\varphi_j}$ form a basis for the scheme output Hilbert space.
The total probability that measurement outcome $m_i$ occurs is then given by
$P(m_i)=\sum_jP(m_i|\varphi_j)P(\varphi_j)$. This definition is consistent with a POVM $\{\hat{F}_i\}$, defined by $\hat{F}_i = \sum_jP(m_i|\varphi_j)\ket{\varphi_j}\bra{\varphi_j}$ so that the probability of measurement outcome $m_i$ is $P(m_i)=\text{Tr}(\hat{F}_i\hat{\rho})$ where $\hat{\rho}=\sum_{j,k}\rho_{jk}\ket{\varphi_j}\bra{\varphi_k}$ and $\rho_{jj}=P(\varphi_j)$. Since by definition $\sum_iP(m_i|\varphi_j)=1$, we have $\sum_i \hat{F}_i=\hat{I}$. By describing the measurement using conditional probabilities directly, we avoid the density operator formalism, which saves computational resources.

Summing $P(m_i)$ over all post-selected measurements and averaging over each Bell state input
gives the total probability of a positive measurement, $p_t+p_f$. The true success rate $p_t$ is
determined in the same way with the exception that the sum is limited to unambiguous output states only, as determined using $\eta_i=1$. The fidelity is then determined by the ratio of these sums.

To calculate the conditional probabilities $P(m_i|\varphi_j)$, each mode of each $\ket{\varphi_j}$ of the output Hilbert space was first passed through a
beam splitter with weight $\sqrt{\eta_d}$ for sub-unity detector efficiency $\eta_d$. This
transformation is given by Eq. (\ref{BSloss}) where $\eta_i$ is replaced by
$\eta_d$. By modeling detector inefficiency using a beam-splitter model, we can consider the parameter $\eta_d$ as an effective efficiency/transmission parameter that implicitly accounts for any photon loss experienced between the scheme and the detectors, such as coupling loss.

We then computed the probabilities associated with each outcome.
At this point the loss modes were safely discarded. To include dark counts, we then allowed each mode to gain one photon with probability $\xi$ or remain unchanged with
probability $1-\xi$.  Here we assume that $\xi\ll 1$ so that the probability of multiple dark counts occurring simultaneously in a single mode is negligible.

If detectors other than PNRDs were being analyzed, each measurement outcome was
modified according to detector model, either by eliminating photon count
information, eliminating the late mode if the early mode is detected, or both.
This can be done only at the level of the probabilities, otherwise
undesired interference can occur and lead to incorrect results. This method
allowed us to simultaneously obtain probabilities $P(m_i|\varphi_j)$ and
identify the subset of relevant measurement outcomes.

\subsection{Detector array analysis}
\label{ssec:metharray}

We now proceed to describe the method we used to calculate analytic expressions
for success rates using arrays of BDs.
The BD-type models that we considered can be assumed to perform ideally if a
single photon is incident when ignoring detector inefficiencies and dark counts.
Therefore by splitting output modes using beam splitters we can on average
reduce the multi-photon detection events that occur and recapture success rate lost by
non-ideal detector types. The method we outline here to determine success rates
is similar to the analysis method proposed by Kok and Braunstein
\cite{kok2001}.

We first give an example for the standard scheme and then extend this to include
the enhanced scheme. To compute the success rate with increasing numbers of
detectors, it is beneficial to first categorize output states based on the
exponents of their photon creation operators. For the standard scheme the two
possible categories are $\{1,1\}$ and $\{2\}$, which for example would include states
such as $\hat{a}^{\dagger}_1\hat{b}^{\dagger}_1$ and $\hat{a}^{\dagger 2}_1$,
respectively.

For the beam splitter detector array model, we always assume that one input mode
of the beam splitter is the vacuum state with the other input being an output
mode of the scheme. In this case, no interference can occur and so we can treat
the array problem using classical statistics.

Let $P\{1,1\}$ and $P\{2\}$ denote the probabilities that an output of type $\{1,1\}$
and $\{2\}$ occur, respectively. These probabilities are computed assuming the input state is an equal distribution of each Bell state so that $P$ represents the average probability over the four Bell states. Similarly, let
$P_t\{1,1\}$ and $P_t\{2\}$ denote the probabilities that output states of those
types also cause collapse into a Bell state, hence give rise to true-positive
events.
In this sense \mbox{$P\{1,1\}+P\{2\} = 1$} and \mbox{$P_t\{1,1\}+P_t\{2\} =
\frac{1}{2}$}.
Although these sums always remain constant, in general the probabilities $P$ and
$P_t$ can change with increasing detector number per array, denoted as $N$. In
particular, since adding beam splitters can only ever convert type $\{2\}$ into type
$\{1,1\}$ states, probability must always flow from $P\{2\}$ to $P\{1, 1\}$ and from
$P_t\{2\}$ to $P_t\{1,1\}$.

To simplify notation, we introduce the probability vectors $\vec{P}
=(P\{1,1\}, P\{2\})^\text{T}$ and $\vec{P}_t =(P_t\{1,1\},P_t\{2\})^\text{T}$. For
example, the standard scheme with BDs gives vectors
$\vec{P}=(\frac{1}{2},\frac{1}{2})^\text{T}$ and
$\vec{P}_t=(\frac{1}{2},0)^\text{T}$ since all outputs of $\ket{\psi^\pm}$ are
of type $\{1,1\}$ and unambiguous, whereas outputs of $\ket{\phi^\pm}$ are of type
$\{2\}$ and ambiguous (see \mbox{Eq. (\ref{eq:outputs})}).

The linear transformation that governs the probability flow for the standard
scheme after $\log_2(N)$ beam splitters is given in the $\{\{1,1\},\{2\}\}$ basis as
\begin{equation}
A(N)=\begin{pmatrix}
1& \frac{1}{2}\\
0& \frac{1}{2}
\end{pmatrix}^{\log_2(N)} =
\begin{pmatrix}
1 & \frac{N-1}{N}\\
0 & \frac{1}{N}
\end{pmatrix}\;,
\end{equation}
\noindent
which can be obtained by analyzing the beam splitter transformation
$\hat{c}^{\dagger}\rightarrow (\hat{c}^{\dagger}_1 +
i\hat{c}^{\dagger}_2)/\sqrt{2}$ applied to each mode of archetype states
representing $\{1,1\}$ and $\{2\}$. Since BDs can only discriminate
terms of type $\{1,1\}$, the probability $P_t\{1,1\}$ after
applying $A(N)$ is the true success rate. For example, for $\vec{P}_t=(\frac{1}{2},0)^\text{T}$, we have that the true success rate \mbox{$p_t(N)=(A(N)\vec{P}_t)_{\{1,1\}} =
\frac{1}{2}$} is constant and equal to the scheme maximum rate, as expected for
the standard scheme with BDs.

To extend this idea to the enhanced scheme, we considered probabilities
associated with five categories, namely $\{1,1,1,1\}$, $\{1,1,2\}$, $\{1,3\}$,
$\{2,2\}$, and $\{4\}$. The transformation $A(N)$ governing the probability flow in the
$\{\{1,1,1,1\}, \{1,1,2\}, \{1,3\}, \{2,2\}, \{4\}\}$ basis is given by
\begin{equation}
\begin{aligned}
\hspace{-3mm}
\left(
\begin{array}{ccccc}
 1 & \frac{N-1}{N} & \frac{(N-2) (N-1)}{N^2} &
 \frac{(N-1)^2}{N^2} & \frac{(N-3) (N-2) (N-1)}{N^3} \\
 0 & \frac{1}{N} & \frac{3 (N-1)}{N^2} &
 \frac{2 (N-1)}{N^2} & \frac{6 (N-2) (N-1)}{N^3} \\
 0 & 0 & \frac{1}{N^2} & 0 & \frac{4 (N-1)}{N^3} \\
 0 & 0 & 0 & \frac{1}{N^2} & \frac{3 (N-1)}{N^3} \\
 0 & 0 & 0 & 0 & \frac{1}{N^3} \\
\end{array}
\right)\!\!.\!\!
\end{aligned}
\end{equation}
\noindent
The true success rate is then given by $(A(N)\vec{P}_t)_{\{1,1,1,1\}}$ where
$\vec{P}_t$ is obtained from the enhanced scheme outputs.

If photon loss is taken into account, the number of categories for both schemes
increases. For the standard scheme we must consider the additional two
categories: $\{0\}$ and $\{1\}$. Likewise, the enhanced scheme has seven
additional categories: $\{0\}$, $\{1\}$, $\{1,1\}$, $\{1,1,1\}$, $\{1,2\}$, $\{2\}$, and $\{3\}$.

\vspace{-2mm}
\section{Results}
\label{sec:results}

\subsection{Scheme efficiency and fidelity}
\label{ssec:rates}

In this section, we report success rates and fidelity for both schemes. For the enhanced
scheme, we assume that the auxiliary state is produced by a single pair source that is subject to loss, representing pair generation inefficiency or propagation/coupling loss.

\subsubsection{Numerical results}

Firstly, without including loss, inefficiencies, or dark counts, we found that each detector type provides enough information to obtain a non-zero success rate with unity fidelity [TABLE \ref{contingency}]. For BD-type models, the optimal measurements to post select are unambiguous measurements that indicate each photon hit a different detector. The probability of this occurring was found to be less in the enhanced scheme than in the standard scheme. Thus, although the enhanced scheme outperformed the standard scheme when PNRD-type models were used, it was inferior to the standard scheme for BD-type models.

\begin{table}
\caption{
A comparison of the maximum success rates $p_{t,\text{max}}$ obtainable with unity fidelity for four different detector models and two different schemes. These success rates are computed under the ideal case; with unity detector efficiency and photon transmission $\eta_d=\eta_i=\eta_a=1$, and zero dark count probability $\xi=0$. Note that the slow-prefixed detector models are relevant to time-bin encoded applications only.
}
\vspace{2mm}
\label{contingency}
\begin{tabular}{c|cccc}
Success Rate     &	PNRD & BD & Slow PNRD & Slow BD           \\ \hline\hline
Standard Scheme 	& 1/2 & 1/2  & 1/2 & 1/4  \\
Enhanced Scheme 	& 3/4 & 3/16 & 39/64 & 1/16
\end{tabular}
\end{table}

For realistic detector parameters \cite{hadfield2009,eisaman2011}, the dark
count probability can be between $10^{-4}$ and $10^{-8}$. For these probabilities,
dark counts have negligible impact on success rates and fidelity for both Bell
analyzers unless detector efficiency is very small ($\eta_d < 0.01$). For the purposes of our numerics, we select a dark count probability of $\xi=10^{-5}$.

\begin{figure}
\hspace{-2mm}
\includegraphics[width=8.7cm]{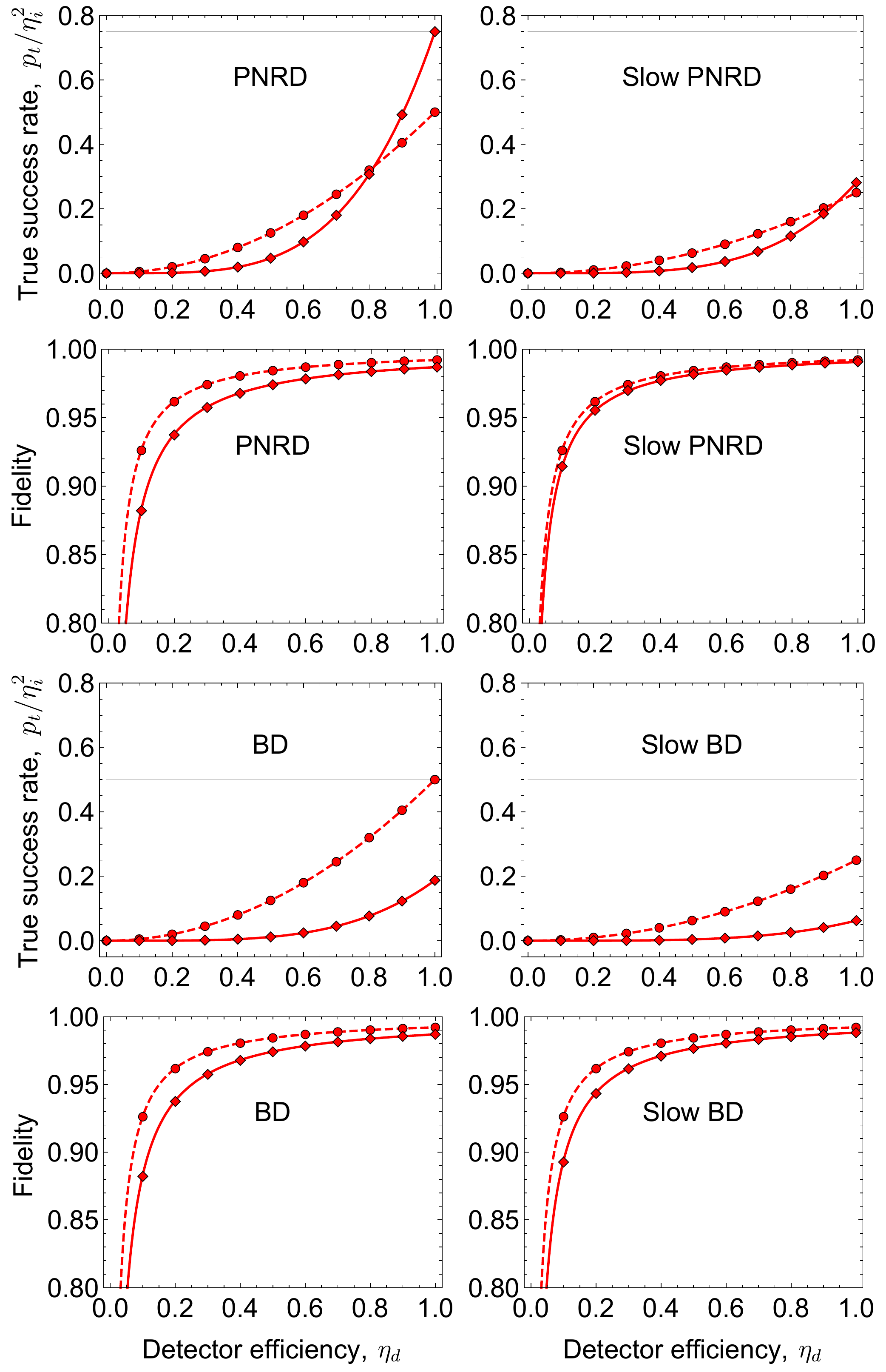}
\caption{
(Color online) The effect of detector efficiency on the true
success rate and fidelity for both the standard and enhanced Bell-state analyzers. Dashed and solid curves represent analytic approximations to the standard and enhanced scheme solutions, respectively. Circle and diamond points represent numerically exact values for the standard and enhanced schemes, respectively. All plots are calculated with a dark count probability of $\xi=10^{-5}$, an auxiliary state transmission of $\eta_a=1.0$, and an input state transmission of $\eta_i=0.01$. The success rates are plotted after dividing out the scheme-independent quantity $\eta_i^2$.  The slow detector models are relevant to time-bin encoded applications only.}
\label{fig:eff}
\end{figure}

In repeater schemes, Bell-state analyzers are used to measure photons that must first
travel over long distances \cite{krovi2015}. This can cause photon losses
around 99\% before Bell-state analysis, effectively enhancing the infidelity caused by dark counts. When including input state loss, acceptable measurement fidelity can only be obtained when $\xi\ll\eta_i,\eta_d$. For this reason, we only explore the regime where first-order dark counts dominate the false-positive rate. For the purposes of our numerics, we select an input state transmission of $\eta_i=0.01$, which is a realistic value in the context of quantum repeaters.  Since a successful Bell-state measurement only occurs when both input photons arrive at the analyzer, the true success rate
scales as $p_t\propto \eta_i^2$. This scaling is independent of the analyzer
scheme and is divided out for a more direct scheme comparison \mbox{[FIG.
\ref{fig:eff}]}.

\begin{figure}
\hspace{-2mm}
\includegraphics[width=8.7cm]{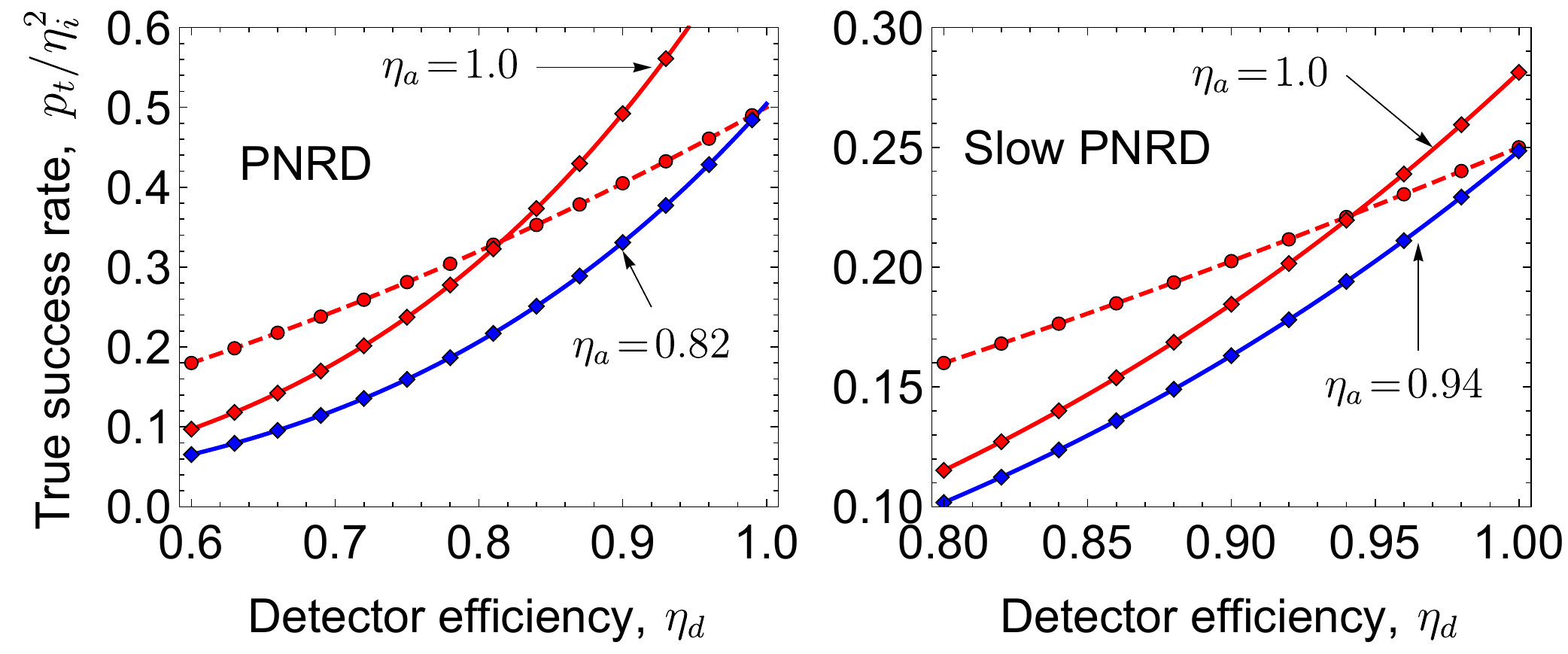}
\caption{
(Color online) A closer view of the success rate intersection points between the standard and enhanced schemes for PNRDs and slow PNRDs for different auxiliary state transmission values $\eta_a$. Dashed and solid curves represent analytic approximations to the standard and enhanced scheme solutions, respectively. Circle and diamond points represent numerically exact values for the standard and enhanced schemes, respectively. All numerical points are calculated with a dark count probability of $\xi=10^{-5}$ and an input transmission of $\eta_i=0.01$. The success rates are plotted after dividing out the scheme-independent quantity $\eta_i^2$. The fidelity plots for different $\eta_a$ are not shown since, for $\eta_a\gg\eta_i$, they are nearly indistinguishable from those shown in FIG. \ref{fig:eff}. The slow PNRD detector model is relevant to time-bin encoded applications only.}
\label{fig:eff2}
\end{figure}

Under dark count and loss conditions, our numerical results showed that the enhanced scheme with \mbox{PNRDs} provided higher
success rates than the standard scheme when $\eta_d>0.82$, assuming $\eta_a=1.0$, or when $\eta_a>0.82$, assuming $\eta_d=1.0$ [FIG. \ref{fig:eff2}]. For slow PNRDs, the enhanced scheme can be superior when $\eta_d>0.94$, assuming $\eta_a=1.0$, or when $\eta_a>0.94$, assuming $\eta_d=1.0$. For BD-types, the enhanced scheme showed
no advantage over the standard scheme for all efficiency values. In addition,
the standard scheme was found to have better fidelity scaling with detector efficiency in
all cases.

It is important to note that all but the slow PNRD cases approach the scheme
maximum rates given in \mbox{TABLE \ref{contingency}} as $\eta_d\rightarrow 1$.
We explain this discrepancy now. For the detector models labeled PNRD, BD, and slow BD, the ideal post-selected measurements do not become ambiguous when only loss and detector inefficiency are introduced. For these three detector types, it is possible to always identify and reject measurements that indicate fewer photons than expected. For PNRDs, this is possible due to ideal number-resolving capabilities. For BD-types, this is possible because the ideal post-selected measurements all already indicate photons arriving at different detectors. Therefore false-positive events cannot occur unless a false state has lost photons that are replaced by an equal number of dark counts during detection, which could then result in a positive measurement. Since dark counts are considered rare, the fidelity is already very good when detector efficiency is large for PNRD, BD, and slow BD models and so we need not reject additional measurements to improve fidelity.

The slow PNRDs main advantage over the slow BD is its ability to distinguish
states based on the photon count in the early mode. As shown in section
\ref{ssec:quality} with the standard scheme, the slow PNRD allowed for a success rate of 50\%, as opposed to the 25\% maximum using a slow BD, by
distinguishing $\ket{\psi^+}$ from $\ket{\phi^\pm}$ based only photon count information.
As a consequence, when using slow PNRDs, the fidelity is extremely susceptible to detector
inefficiency and photon loss, even without the presence of dark counts. In the standard scheme, for example, detector inefficiency or photon loss
can cause the two-photon states from $\ket{\phi^\pm}$ to often appear as a
single early photon, which are then falsely post-selected as indicating the
$\ket{\psi^+}$ Bell state. Thus, when using photon count information from slow PNRDs, photon loss or detector inefficiency can cause false-positive events even in the absence of dark counts, if these specific types of measurements are post-selected.

In order to maintain acceptable fidelity when using slow PNRDs, it is necessary to further reject measurements that become ambiguous
when losses are introduced. This is accomplished by only accepting measurements that preserve the expected total photon count. For the standard scheme, this is equivalent to completely ignoring photon count information, effectively treating the slow PNRD as a slow BD, giving a maximum success rate of 1/4. For the
enhanced scheme, this resulted in an improved success
rate over the corresponding slow BD case with a maximum success rate of 9/32. These two values are the success rates for $\eta_d=1.0$ illustrated in the slow PNRD case in FIG. \ref{fig:eff}.

\subsubsection{Analytic approximations}
\vspace{-2mm}

Once the maximum success rates $p_{t,\text{max}}$ are determined for a given detector model and scheme, understanding the effect of losses and inefficiencies on the success rate is uncomplicated. The vast majority of true-positive measurements occur when detectors operate correctly and no photons are lost. For the standard scheme, the two possible loss channels are input loss and detector inefficiency, represented by parameters $\eta_i$ and $\eta_d$, respectively. Since the standard scheme utilizes two incident photons, two detectors, and has four modes that experience dark counts, the total success rate is given by $p_t = \eta_d^2\eta_i^2(1-\xi)^4p_{t,\text{max}}+p_\xi$, where $p_{t,\text{max}}$ is $1/2$ in the PNRD and BD cases and $1/4$ for the slow-type models. The small correction $p_\xi$ is a positive contribution to the success rate caused by true-positive cases resulting from inefficiencies and dark counts effectively canceling each other, leaving the measurement conclusion unaffected. For dark count probability $\xi\ll 1$, we have $(1-\xi)^4\simeq 1$ and $p_\xi\simeq 0$ leaving $p_t\simeq \eta_d^2\eta_i^2p_{t,\text{max}}$.

The enhanced scheme can be analyzed in a similar manner. In addition to the two loss channels listed above, the enhanced scheme can suffer from auxiliary state photon loss, parameterized by $\eta_a$. When the enhanced scheme has no auxiliary input, the scheme resorts to behaving like the standard scheme, with a maximum success rate of $50\%$ \cite{grice2011}. Hence, when $\eta_a\neq 1$, some success rate could be recovered by post-selecting some measurements where both auxiliary photons are lost, in addition to post-selecting measurements where the auxiliary state is unaffected. Unfortunately, if this is done, false-positive measurements can arise from input state photon loss directly, similar to the slow PNRD case discussed previously. With large input loss, the reduction in fidelity caused by including measurements resulting from a vacuum auxiliary state will not justify the gain in success rate, so we only consider post-selecting $\eta_a^2$ terms. Since the enhanced scheme utilizes four incident photons, four detectors, and has eight modes that can experience a dark count, the total success rate is given by $p_t=\eta_d^4\eta_a^2\eta_i^2(1-\xi)^8p_{t,\text{max}}+p_\xi$, where $p_{t,\text{max}}$ is $3/4$, $3/16$, $9/32$, and $1/16$ for the PNRD, BD, slow PNRD, and slow BD cases, respectively. Again, for $\xi\ll 1$ we have $(1-\xi)^8\simeq 1$ and $p_\xi\simeq 0$ so that $p_t\simeq\eta_d^4\eta_a^2\eta_i^2p_{t,\text{max}}$. These zero-order terms in $\xi$ well-approximate the numerical results in FIG. \ref{fig:eff} and FIG. \ref{fig:eff2}.

With these results, it is straightforward to predict the bounds where the enhanced scheme is superior to the standard scheme. For PNRDs we have \mbox{$3\eta_i^2\eta_d^4\eta_a^2/4>\eta_i^2\eta_d^2/2$} implying \mbox{$\eta_d\eta_a > \sqrt{2/3}\simeq0.82$}. For slow PNRDs we have \mbox{$9\eta_i^2\eta_d^4\eta_a^2/32>\eta_i^2\eta_d^2/4$} implying \mbox{$\eta_d\eta_a>2\sqrt{2}/3\simeq0.94$}. Both of these expressions agree with our numerical results. These results imply that the enhanced scheme is superior to the standard scheme in terms of success rate if efficient detectors with at least some number resolving capabilities are utilized and bright pair sources are available to produce the auxiliary state.

Understanding the effect of loss and dark counts on fidelity can be done in a similar manner. Since only measurements that preserve the expected photon count are post-selected, all false-positive measurements must arise from at least one dark count event. In this work we assume $\xi\ll \eta_i\ll 1$ so that only terms first-order in dark count probability dominate the false-positive rate. This assumption is justified in any regime where the fidelity is acceptable. It also provides an upper bound on the fidelity even in the regime $\xi\simeq\eta_i$. In addition, if we further assume that $\eta_i\ll\eta_d,\eta_a$, the dominant term will also correspond to the case where a single input photon is lost before arriving at the analyzer, and no other photons are lost. This assumption is reasonable for a repeater scheme, where input state transmission is poor.

Knowing that, in the regime $\xi\ll\eta_i\ll\eta_d,\eta_a$, the false-positive rate is dominated by one-photon input loss and single dark count events, the false-positive rates are of the form \mbox{$p_f\simeq C_1\eta_d\eta_i(1-\eta_i)\xi(1-\xi)^3$} for the standard scheme and $p_f\simeq C_2\eta_a^2\eta_d^3\eta_i(1-\eta_i)\xi(1-\xi)^7$ for the enhanced scheme. The constant prefactors, $C_1$ and $C_2$, depend on detector model and can be solved by directly counting the probability that single-photon inputs are post-selected, given the output states and post-selected measurements of each scheme. For example, for the standard scheme with PNRD detectors, the post-selected measurements indicate either $\hat{a}^\dagger_1\hat{b}^\dagger_2\ket{0}$, $\hat{a}^\dagger_2\hat{b}^\dagger_1\ket{0}$, $\hat{a}^\dagger_1\hat{b}^\dagger_1\ket{0}$, or $\hat{a}^\dagger_2\hat{b}^\dagger_2\ket{0}$ (see section \ref{ssec:quality}). The probability of obtaining a single photon output is $2\eta_i(1-\eta_i)$ for each mode. This single photon can only become post-selected if a dark count occurs in two of three possible modes. For example, $\hat{a}^\dagger_1\ket{0}$ would require a dark count in either $\hat{b}^\dagger_2$ or $\hat{b}^\dagger_1$ to give rise to a false positive measurement. Hence, in this case, the prefactor is $C_1=4$. Likewise, for the slow BD case, only measurements indicating $\hat{a}^\dagger_1\hat{b}^\dagger_2\ket{0}$ or $\hat{a}^\dagger_2\hat{b}^\dagger_1\ket{0}$ are post-selected. Here, a single photon can only become post-selected if a dark count occurs in one of three possible modes. Hence the prefactor is $C_1=2$, where the 2 arises from the probability of obtaining a single photon output.

The prefactors $C_1$ for the standard scheme were found to be 4 for PNRD and BD models and 2 for the slow detectors. The enhanced scheme prefactors $C_2$ were found to be 10, 5/2, 8/3, and 3/4 for PNRD, BD, slow PNRD, and slow BD models, respectively. These false positive rates were used along with the analytic success rate approximations to determine the fidelity. The resulting fidelity well-approximates the numerically exact results in FIG. \ref{fig:eff} and FIG. \ref{fig:eff2}.

An interesting consequence of the result that both $p_t\propto \eta_a^2$ and $p_f\propto\eta_a^2$ is that the fidelity, $p_t/(p_t+p_f)$, is independent of the auxiliary state photon loss rate. In addition, $p_f\propto \xi$ for $\xi\ll\eta_i\ll 1$ indicates that any difference in fidelity between the standard and enhanced schemes can be made negligible with a small enough detector noise. Decreasing the dark count probability below $10^{-6}$ would show significant improvements to fidelity with inefficient detectors for both schemes. 

\vspace{-2mm}
\subsection{Detector arrays}
\label{ssec:detarr}
\vspace{-2mm}

In this section, we report success rates and fidelity for both schemes assuming the detectors are replaced by arrays of BD or slow BD detectors. For the enhanced
scheme, we still assume that the auxiliary state is produced by a single pair source that is subject to loss, representing pair generation inefficiency or propagation/coupling loss.

\subsubsection{Numerical results}

\begin{figure}[t!]
\hspace{-3mm}
\includegraphics[width=8.8cm]{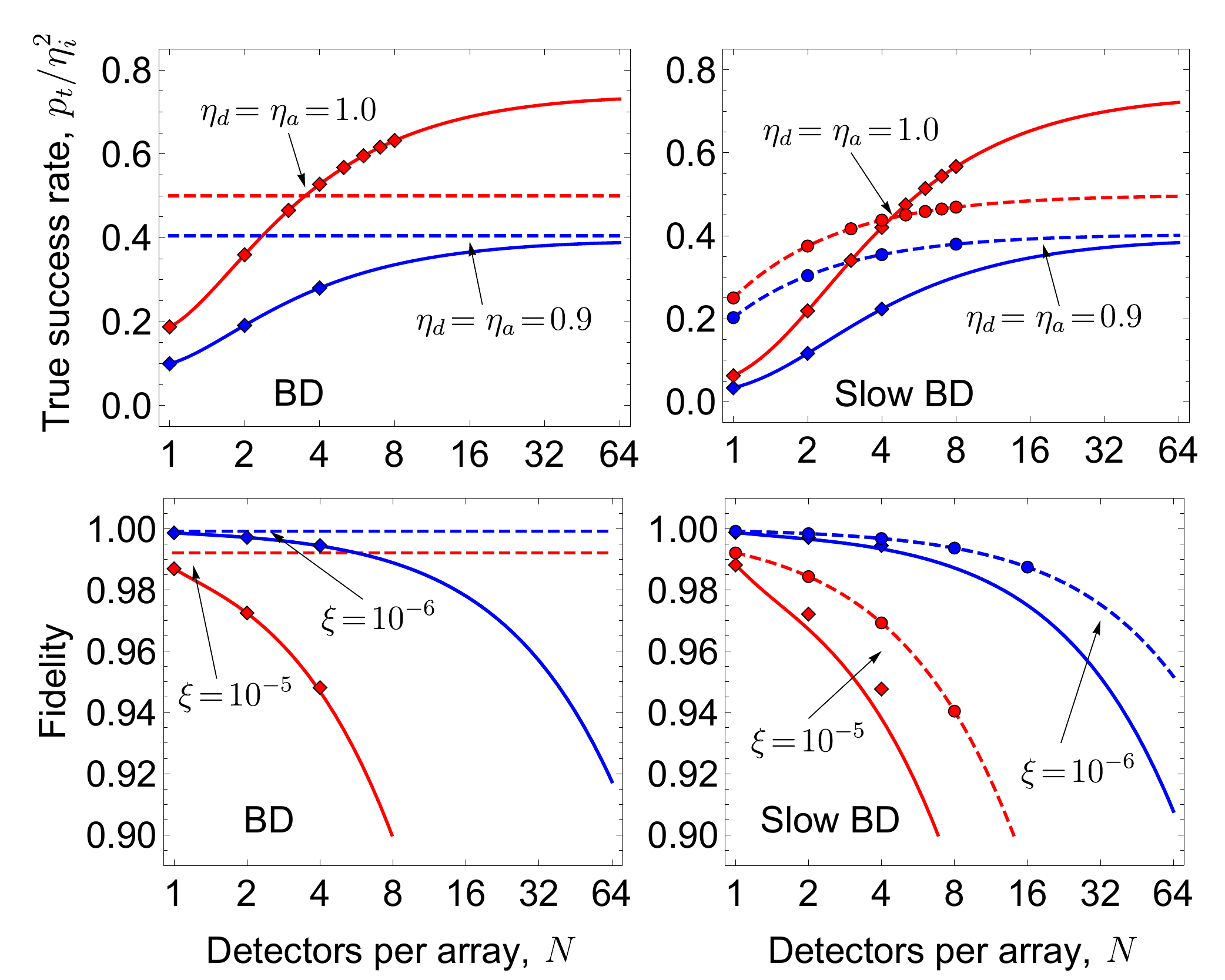}
\caption{
(Color online) Increase in true success rates using arrays of imperfect
detectors to approximate photon number resolving detectors. Dashed and solid curves represent analytic approximations to the standard and enhanced scheme solutions, respectively. Circle and diamond points represent numerically exact values computed for computationally reasonable $N$. The fidelity curves for the enhanced scheme (solid) are lower bound
approximations obtained from upper bound
false-positive rates. The fidelity curves and data points were computed using input state transmission $\eta_i=0.01$, detector efficiency $\eta_d=1.0$, two different dark count probabilities, $\xi=10^{-5}$ or $\xi=10^{-6}$, and an auxiliary state transmission $\eta_a=1.0$ when relevant. The success rates are plotted after dividing out the scheme-independent quantity $\eta_i^2$. The dashed lines for the BD case show the $N=1$ values for the standard scheme since arrays of BDs cannot improve the standard scheme success rate. The slow BD detector model is relevant to time-bin encoded applications only.
}
\label{fig:arrays}
\vspace{-2mm}
\end{figure}

Firstly, with high detector efficiency and a very bright auxiliary pair source, we found that arrays of BD or slow BD types can allow the enhanced scheme success rate to exceed that of the standard scheme [FIG. \ref{fig:arrays}]. Note that the success rate for the standard scheme using BDs is already optimal without using detector arrays and adding arrays of BDs to the standard scheme would only decrease the fidelity. For this reason, we compare the enhanced scheme BD case to the optimal $N=1$ values for the standard scheme BD case. In this case, the enhanced scheme becomes superior to the standard scheme when four detectors are used per output (16 total).

For slow BDs, using detector arrays improves the success rate for both the standard and enhanced schemes. However, due to a larger maximum success rate, the enhanced scheme can surpass the standard scheme with arrays of at least five detectors per output (20 total). Note that a uniform
array of size other than a power of two cannot be implemented using just 50/50
beam splitters but could be implemented using multi-port beam splitters or
uneven (non 50/50) beam splitters. This model also gives upper bound rates for
other spatial multiplexing methods such as beam-shaping optics \cite{jiang2007}.

Adding arrays of non-number-resolving detectors will only ever recover success rate up to the success rate achievable with PNRDs. For the enhanced scheme to become superior to the standard scheme using either BDs or slow BDs with increasing array size, we still require $\eta_a\eta_d>0.82$ (see section \ref{ssec:rates}). Decreasing the detector efficiency or the auxiliary state transmission decreases the success rate, as expected, and moves the intersection point between the standard and enhanced scheme success rate to larger array sizes.

When using large arrays, the detector dark counts have a large impact on the fidelity when input state photon loss is high, even if detector efficiency is high [FIG. \ref{fig:arrays}]. In the array size regime where the enhanced scheme overtakes the standard scheme in success rate ($N>4$), a dark count probability of \mbox{$\xi\leq 10^{-6}$} is required for acceptable fidelity. Numerically computing the fidelity and success rate becomes computationally difficult for array sizes $N\geq 8$ for the enhanced scheme. This is partially due to the large Hilbert space size, which is 814385 for $N=8$ when including photon loss, but also due to the very large number of measurement outcomes, which could be up to $\sum_{i=0}^5\binom{64}{i}=8303633$ in the BD case for 4 photons and one dark count in 64 modes. Monte Carlo methods might prove a better method to compute numerical values for larger array sizes \cite{jiang2007}. In the following section, we determine analytic approximations that give insight into the impact of imperfections on the success rate and fidelity for larger array sizes.

\subsubsection{Analytic approximations}

To make analytic predictions for the success rate and the fidelity, we use the methods described in section \ref{ssec:metharray}. Recall that, as an example, in section \ref{ssec:metharray} we showed that the standard scheme with BDs is associated with probability vectors $\vec{P}=(\frac{1}{2},\frac{1}{2})^\text{T}$ and
$\vec{P}_t=(\frac{1}{2},0)^\text{T}$ in the \{\{1,1\}, \{2\}\} basis. This resulted
in the constant success rate $p_t(N)=1/2$. Now, we use the same reasoning to solve the standard scheme with slow BDs as well as the enhanced scheme with BDs and slow BDs.

To determine the success rate expression for the standard scheme with slow BDs, we first found the
effective probability vectors for the slow BD model. These vectors are different
from the regular BD case and, at the level of the probabilities, can be computed from the
scheme outputs by combining early and late time-bin modes of the same spatial
mode if the early mode is occupied. Notice that in this case we do not wish to
eliminate the photons in the late mode since for this method to work correctly
the total photon count must be preserved. That is, we need only capture the
behavior of the `slow' flaw since the non-number-resolving flaw of the slow BD
is automatically captured by only allowing probability of $\{1,1\}$ types to
contribute to success rate. Performing this modification to the output states of
the standard scheme gives the vectors $\vec{P} =
(\frac{1}{4},\frac{3}{4})^\text{T}$ and $\vec{P}_t =
(\frac{1}{4},\frac{1}{4})^\text{T}$. By applying the array transformation $A(N)$
to $\vec{P}_t$ we get the success rate
\begin{equation}
p_t(N) = (A(N)\vec{P}_t)_{\{1,1\}} = \frac{1}{2} -\frac{1}{4N}\;.
\end{equation}

The enhanced scheme output gives probability vectors
describing the initial distribution over the categories in the basis
\hbox{\{\{1,1,1,1\}, \{1,1,2\}, \{1,3\}, \{2,2\}, \{4\}\}} as
\begin{equation}
\begin{aligned}
\hspace{-2.6mm}
\vec{P} = \hspace{-1mm}\left(\frac{1}{4}, \frac{5}{16}, \frac{3}{16},
\frac{5}{32}, \frac{3}{32}\right)^\text{\!\!\!T}\hspace{1mm}
\vec{P}_t = \hspace{-1mm}\left(\frac{3}{16}, \frac{5}{16}, \frac{3}{16},
\frac{1}{16}, 0\!\right)^\text{\!\!\!T}\!
\end{aligned}
\end{equation}
\noindent
for the BD model. Applying the transformation $A(N)$ for the enhanced scheme
gives success rate
\begin{equation}
p_t(N) = (A(N)\vec{P}_t)_{\{1,1,1,1\}} = \frac{3}{4} -\frac{1}{N} +
\frac{7}{16N^2}\;.
\end{equation}
\noindent
Likewise, for the slow BD model we have
\begin{equation}
\begin{aligned}
\hspace{-3mm}
\vec{P} = \hspace{-1mm}\left(\frac{1}{8}, \frac{1}{4}, \frac{1}{4},
\frac{7}{32}, \frac{5}{32}\right)^\text{\!\!\!T}\hspace{2mm}
\vec{P}_t = \hspace{-1mm}\left(\frac{1}{16}, \frac{1}{4}, \frac{1}{4},
\frac{1}{8}, \frac{1}{16}\right)^\text{\!\!\!T}\;,
\end{aligned}
\end{equation}
\noindent
which results in
\begin{equation}
p_t(N) = \frac{3}{4} -\frac{13}{8N} + \frac{21}{16N^2} - \frac{3}{8N^3}\;.
\end{equation}
Taking $N\rightarrow\infty$ we see that $p_t(N)$ approaches the scheme maximum
rates in all cases, as expected.

To include dark counts, detector inefficiency, and photon loss, we again assumed that
the vast majority of true-positive events occur when detectors operate correctly
and all photons arrive at the analyzer. This gives $p_t \simeq
\eta_d^2\eta_i^2(1-\xi)^{4N}p_t(N)$ for the standard scheme and $p_t \simeq
\eta_d^4\eta_a^2\eta_i^2(1-\xi)^{8N}p_t(N)$ for the enhanced scheme. The term $4N$ and $8N$ are the total number of modes after
applying an $N$-sized beam splitter model for the standard and enhanced schemes, respectively.

To estimate the false-positive rate while including imperfections, we use the same arguments as in the $N=1$ case in section \ref{ssec:rates}. However, for detector arrays, the prefactors $C_1$ and $C_2$ become dependent on $N$ in a nontrivial way. For the standard scheme, the prefactor is uncomplicated because most false-positive measurements arise from states of type $\{1\}$, and $P\{1\}=2\eta_i(1-\eta_i)$ is independent of $N$. For slow BDs, by inspection, the number of false-positive combinations was determined to be $(2N-1)$, and so $C_1(N)=2(2N-1)$. Hence the false-positive rate for the standard scheme with arrays of $N$ slow BDs is approximately
\begin{equation}
p_f \simeq 2(2N-1)\eta_d\eta_i(1-\eta_i)\xi(1-\xi)^{4N-1}.
\end{equation}

The enhanced scheme false-positive rates are more difficult to determine because most false-positive measurements arise from states of type $\{1,1,1\}$, and $P\{1,1,1\}$ increases with $N$ due to contribution from states of type $\{1,2\}$ and $\{3\}$ splitting. Using $A(N)$ with input state and auxiliary state loss considerations, the $N$-dependent $P\{1,1,1\}$ is
\begin{equation}
\label{ESBD111}
P\{1,1,1\} \simeq \eta_a^2\eta_i(1-\eta_i)\left(2-\frac{2}{N}+\frac{3}{4N^2}\right)
\end{equation}
for the BD model, assuming $\eta_i\ll\eta_a$. Likewise, for the slow BD model we have
\begin{equation}
\label{ESslowBD111}
P\{1,1,1\} \simeq \eta_a^2\eta_i(1-\eta_i)\left(2-\frac{5}{2N}+\frac{1}{N^2}\right)\!.
\end{equation}

An approximate upper bound for the false-positive rate is given by the probability that a state of type $\{1,1,1\}$ is measured to be of type $\{1,1,1,1\}$. For BD detectors, $\{1,1,1\}$-type states have 3 of the $8N$ modes occupied and so $8N-3$ modes could suffer a dark count to become a $\{1,1,1,1\}$ type. This gives an upper bound \mbox{$p_f\leq(8N-3)\eta_d^3\xi(1-\xi)^{8N-1}P\{1,1,1\}$}. This is an upper bound because a fraction of the $\{1,1,1,1\}$ measurements are actually rejected during post-selection. For the $N=1$ BD case, the fraction of rejected states can be directly counted to be 2/3 by inspection of the scheme output states. This fraction was found to have a very slight decreasing trend with increasing $N$, implying 2/3 can provide a more accurate upper bound false-positive rate. Thus the false-positive rate for the enhanced scheme with arrays of $N$ BDs is approximated by
\begin{equation}
p_f\simeq \frac{2}{3}(8N-3)\eta_d^3\xi(1-\xi)^{8N-1}P\{1,1,1\},
\end{equation}
where $P\{1,1,1\}$ is given by Eq. (\ref{ESBD111}). Note that when $N=1$, we recover the result from section \ref{ssec:rates} for BDs. That is, $(2/3)(8N-3)(2-2/N+3/4N^2)=5/2$ for $N=1$. Since the 2/3 factor is an upper bound, the fidelity derived from this false-positive approximation will slightly underestimate the numerical values for fidelity at higher $N$ [FIG. \ref{fig:arrays}].

For slow BD detectors, $\{1,1,1\}$-type states have \mbox{$8N-6$} modes that could suffer a dark count to become a $\{1,1,1,1\}$ type. This is because a dark count must occur in a different \emph{spatial} mode than those already occupied to change the total measured photon count, as a consequence of the slow recovery property. This gives an upper bound \mbox{$p_f\leq(8N-6)\eta_d^3\xi(1-\xi)^{8N-1}P\{1,1,1\}$}. For the $N=1$ slow BD case, the fraction of rejected states can be directly counted to be 3/4, again by inspection of the scheme output states. This fraction was again found to have a decreasing trend, slightly more significant than in the BD case. Thus the false-positive rate for the enhanced scheme with arrays of $N$ slow BDs is approximately
\begin{equation}
p_f\simeq \frac{3}{4}(8N-6)\eta_d^3\xi(1-\xi)^{8N-1}P\{1,1,1\},
\end{equation}
where $P\{1,1,1\}$ is given by Eq. (\ref{ESslowBD111}). With $N=1$, we again recover the result from section \ref{ssec:rates} for slow BDs. That is, $(3/4)(8N-6)(2-5/2N+1/N^2)=3/4$ for $N=1$. The fidelity derived from this false-positive approximation will underestimate the numerical values for fidelity at higher $N$ more so than the corresponding BD fidelity [FIG. \ref{fig:arrays}].

\subsection{Alternative auxiliary source}
\label{ssec:spdc}

We now briefly discuss the case where the enhanced scheme auxiliary state is generated by a SPDC source. In this section we still consider the input state $\ket{\Psi}_{12}$ to be a bi-photon state so that the state being measured is in the span of the Bell basis (see section \ref{ssec:quality}). In this case, $\ket{\Psi}_{12}\ket{\phi^+_n}_{34}$ has constant photon number equal to $2n+2$.
This implies that states with different $n$ are orthogonal and will not interfere when passing
through the analyzer. Hence it suffices to analyze the performance
of the scheme using auxiliary states $\ket{\phi^+_n}_{34}$ separately and
average the resulting rates using the weight function $w^2(n,\tau)$ (see Eq. (\ref{pdctau})).

If the input is a superposition of states with different photon numbers
(such as produced by another SPDC source), then this analysis method fails.
However, the input state need not be a true bi-photon state to test the Bell-state analyzer. Two high-quality single photons could be heralded from weakly-pumped SPDC sources \cite{fasel2004} and then projected
onto a Bell state by the measurement. The efficiency and fidelity of the measurement can be inferred using a decoy state protocol, as if the input was a true bi-photon state \cite{valivarthi2014}.

\begin{figure}
\hspace{-5mm}
\includegraphics[width=8.6cm]{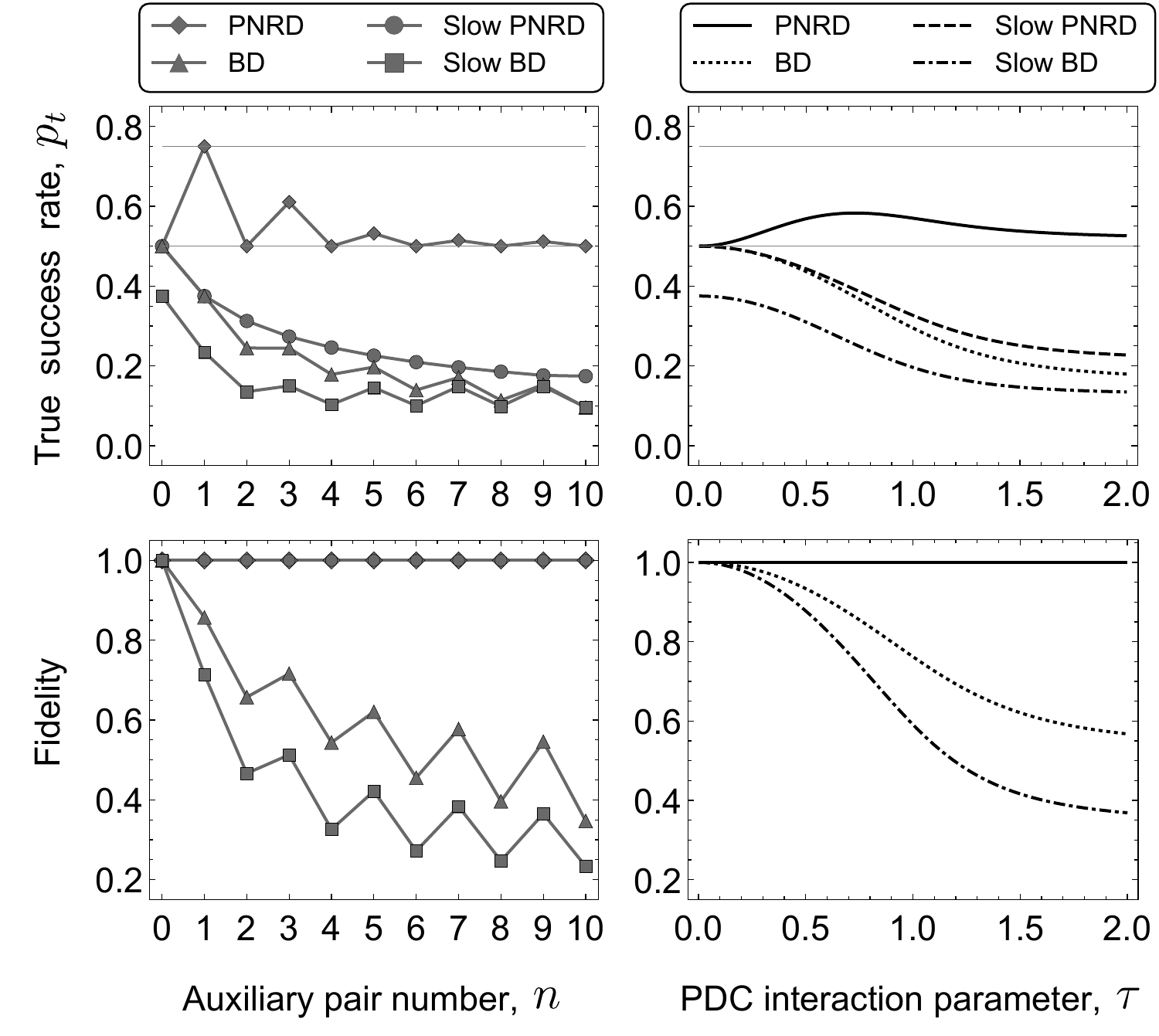}
\caption{
(left) Success rates and fidelity for the enhanced scheme with
different multi-photon auxiliary states from a SPDC source. (right) True
success rates and fidelity, assuming a SPDC-based auxiliary source, after a
weighted average of the true-positive and false-positive rates using weighting
function $w^2(n,\tau)$. Curves on the right correspond to PNRD (solid),
slow PNRD (dashed), BD (dotted), and slow BD (dot-dashed) detector types. All plots were computed in the ideal case; with unity detector efficiency and photon transmission $\eta_d=\eta_i=\eta_a=1$, and zero dark count probability $\xi=0$. The slow-prefixed detector models are relevant to time-bin encoded applications only.
}
\label{fig:pdc}
\end{figure}

To analyze the enhanced scheme using a SPDC auxiliary source, we begin by determining the maximum success rate for each detector type, in the absence of dark counts, detector inefficiencies, and loss rates. To do this, we consider each $\ket{\phi^+_n}$ for $0\leq n \leq 10$ as an auxiliary
state separately. For each $n$, the true-positive measurements were
determined and then combined to form four mutually exclusive sets of
measurements that are used to identify each of the four Bell states. These
combined sets of post-selected measurements were then used to determine the true-positive and false-positive rates for each auxiliary state $\ket{\phi_n^+}$ along with the
associated fidelity. This was done for each detector type, the results of which
are summarized in FIG. \ref{fig:pdc}.

Interestingly, the maximum success rate for the PNRD case remains above 50\%
regardless of the auxiliary pair number. This implies that the enhanced scheme
with a SPDC auxiliary source still has an upper bound rate greater than that of
the ideal standard scheme.
The maximum success rate for the enhanced scheme using a SPDC source is 58\%,
which occurs with an interaction parameter value of $\tau=0.67$, corresponding to a mean photon pair number $\overline{n}=1$ for the auxiliary source.
Values for mean photon pair numbers when generating Bell pairs using SPDC are
generally chosen to be small ($\sim 0.1$) \cite{gisin2002}, since the
multi-photon states are undesired. However, even though the enhanced scheme
operates ideally with a single Bell pair, FIG. \ref{fig:pdc} suggests that the optimal mean photon pair number is much
larger than 0.1 since it is possible to utilize both the empty pulses and the
$n>1$ entangled states generated by SPDC to perform a successful BSM. If the distribution of pair numbers can be
focused nearer to $n=1$, it may be possible to raise the success rate higher
than 58\%. Perhaps this could be achieved by antibunching emission of photon
pairs from the SPDC source using a quantum Zeno blockade \cite{huang2012},
although this could never increase the rate above the 75\% maximum.

Since PNRDs can indicate photon number, each measurement can be attributed
directly to one of the terms in the auxiliary state. For this reason combining
unambiguous measurements from each $n$ results in a set of measurements that
give perfect fidelity for any interaction parameter. Similarly, the partial
number-resolving capability of the slow PNRD model allows a set of unambiguous
measurements to exist, which allows measurements with perfect fidelity, albeit at an inferior rate to the standard scheme. On the other hand,
BD-types cannot indicate photon count and so measurements from one pair number
can be confused with those from another, leading to unavoidable false positive
events.

Computing success rates and fidelity for large auxiliary states is computationally difficult even without including losses, inefficiencies, and detector dark counts. For $n=10$, the computation involves 22 photons, leading to an output Hilbert space of size $N_H=1560780$ (see section \ref{ssec:sizes}). Including all photon losses increases the size to $5852925$. Since the number of measurement outcomes can be even larger than the Hilbert space size when using PNRDs with detector dark counts, we do not attempt numerical computations accounting for these additional imperfections. Instead, we make some observations for small $n$ to make a definitive comparison between the two schemes when using an SPDC auxiliary source.

For the enhanced scheme to be superior to the standard scheme when using PNRDs with an SPDC auxiliary source, it is necessary to accept many measurements that indicate different $n$. As a consequence, the probability of losing two photons can directly contribute to infidelity in the absence of dark counts, similar to the slow PNRD case discussed in section \ref{ssec:rates}. Consider the incident states $\ket{00}_{12}\ket{\phi^+_2}_{34}$ and $\ket{\phi^+}_{12}\ket{\phi^+_1}_{34}$. Due to the symmetry of the apparatus, these two states produce many identical outputs. Unsurprisingly, we found that $\ket{00}_{12}\ket{\phi^+_2}_{34}$ can produce all of the measurement results that are post-selected to indicate a $\ket{\phi^+}_{12}\ket{\phi^+_1}$ input. This implies that $p_f\propto (1-\eta_i)^2$, assuming $\eta_i\ll\eta_d,\eta_a$ and $\xi\ll 1$. Thus for small $\eta_i$, the fidelity is likely also small, unlike previous cases where the false-positive measurements were dominated by single detector dark count cases. For this reason, SPDC sources are not a suitable replacement for the auxiliary source in the enhanced scheme when $\eta_i\ll 1$. This alternative auxiliary source could still be used in a local application with very efficient number resolving detectors and low photon loss rates. In this case, the results shown in FIG. \ref{fig:pdc} become relevant. Regardless, an increase of $0.08$ in maximum success rate over the standard scheme is likely not enough to justify the required increase in apparatus quality and complexity.

\section{Conclusions}
\label{sec:conc}

In this paper, we determined realistic success rates and measurement discrimination fidelities expected when implementing
an enhanced linear-optic Bell measurement scheme using non-ideal detectors and incident states. This enhanced scheme was compared to the standard scheme for implementing a Bell-state analyzer. We described the numeric methods that we used to quantify the quality of the Bell-state analyzers in light of the challenges caused by large Hilbert space sizes when detector arrays and complicated auxiliary states are considered. In addition, we used analytic methods to determine approximate expressions that give insight into the effect of different imperfections on the scheme quality.

When detector inefficiencies, dark counts, input photon losses, and auxiliary photon losses were accounted
for, the enhanced scheme using PNRDs was found to be superior to the standard
scheme when detector and auxiliary source efficiency was high, \mbox{$\eta_d\eta_a>0.82$}. When using single non-PNRDs, the
enhanced scheme showed no advantage over the standard scheme. However, the
enhanced scheme's 75\% maximum rate allowed it to surpass the standard method
when used with efficient non-PNRDs in array configurations.

Both schemes that were analyzed suffered decreases in success rate when used
with time-bin qubits and detectors with long dead time. In this case,
arrays of highly efficient detectors again improved success rates arbitrarily
close to scheme maximums in the large array limit. However, detector arrays are
significantly susceptible to detector dark counts since the chance of misfire
during a detection window is much higher with more active detectors. This effect
is compounded by input loss, causing a reducing fidelity with increasing array sizes. If detectors composing the arrays have dark count probabilities smaller than $10^{-6}$, the measurement fidelity can be made acceptable even with large input loss. 

We analyzed whether a parametric down-conversion source can be used to produce
the auxiliary photon state required for the enhanced scheme. With this
alternative auxiliary source and number-resolving detectors, the enhanced scheme had a maximum success rate of
$58\%$, still greater than that of the standard scheme. However, the measurement discrimination fidelity is very low when input loss is large. Hence, using a SPDC source with the enhanced scheme would only be advantageous for local applications with small loss rates and efficient number-resolving detectors.

Although the regime where this enhanced scheme surpasses the standard scheme is narrow, the advent of accurate, bright pair sources and improved detectors might allow an auxiliary photon assisted Bell-state analyzer to become viable in the near future. Furthermore, the methods and results detailed in this paper may prove useful for the analysis of the performance of other linear optical quantum information processing protocols under realistic conditions.

\section*{Acknowledgments}

This work was supported by the Program for Undergraduate Research Experience
(PURE) at the University of Calgary, Alberta Innovates - Technology Futures
(AITF), and the U.S. Defense Advanced Research Projects Agency (DARPA).
W.T. is a senior fellow of the Canadian Institute for Advanced Research (CIFAR).
We would like to thank Saikat Guha for useful feedback. S.W.
would like to thank Joshua Slater for discussions on detector types.

\end{document}